\documentclass{amsart}
\usepackage{amsmath}
\usepackage{graphicx}
\usepackage{amsthm}
\usepackage{amsfonts}
\usepackage[round]{natbib}

\begin{document}
\title[Logical PPS paradoxes, measurement-disturbance and contextuality]{%
Logical Pre- and Post-Selection paradoxes, measurement-disturbance and
contextuality}
\author{M. S. Leifer}
\email{mleifer@perimeterinstitute.ca}
\author{R. W. Spekkens}
\email{rspekkens@perimeterinstitute.ca}
\address{M. S. Leifer and R. W. Spekkens, Perimeter Institute for Theoretical Physics, 31
Caroline Street North, Waterloo, Ontario, Canada, N2L 2Y5}
\date{October 31, 2004}
\thanks{keywords: pre-selection, post-selection, contextuality, hidden variables,
ABL rule}
\thanks{Pacs: 03.65.Ta}

\begin{abstract}
Many seemingly paradoxical effects are known in the predictions
for outcomes of measurements made on pre- and post-selected
quantum systems. A class of such effects, which we call
\textquotedblleft logical pre- and
post-selection paradoxes\textquotedblright, bear a striking
resemblance to proofs of the Bell-Kochen-Specker theorem, which
suggests that they demonstrate the
contextuality of quantum mechanics. Despite the apparent similarity, we show that such effects can
occur in noncontextual hidden variable theories, provided
measurements are allowed to disturb the values of the hidden
variables.
\end{abstract}

\maketitle

\section{Introduction}

\label{Intro}

The study of quantum systems that are both pre- and post-selected
was initiated by Aharonov, Bergmann and Lebowitz (ABL)
\citep{ABLRule}, and has led to the discovery of many
counter-intuitive results, which we call Pre- and Post-Selection
(PPS) paradoxes\footnote{For a recent review of these results see
\citet{AhaVaidReview}.}.  These results have led to a long debate about the interpretation
of the ABL probability rule\footnote{See \citet{Kastner7} and
references therein.}. 

An undercurrent in this debate has been the
connection between PPS paradoxes and contextuality. \citet{BubBrown} understood the first paper describing a PPS paradox,
that of \citet{Proto3Box}, as a claim to a
novel proof of the contextuality of Hidden Variable Theories
(HVTs), that is, as a version of the Bell-Kochen Specker theorem
\citep{BellKS, KS}, and convincingly disputed this claim. Although
the language of \citet{Proto3Box} does suggest
such a reading\footnote{%
For instance, it is stated that
\textquotedblleft The assumption of Gleason and of Kochen and
Specker [...] is not satisfied by quantum mechanical systems
within the interval between two measurements!\textquotedblright },
in \citet{AADcomment} these authors clarified their
position, stating that it was not their intention to conclude
anything about HVTs. Nonetheless, discussions of PPS paradoxes
continue to make use of a language that suggests implications for
ontology \citep{VaidTS3} and there are explicit claims that PPS
paradoxes are proofs of contextuality \citep{Marchildon}.


In this paper, we show that PPS paradoxes can be explained without
invoking contextuality if one allows that measurement interactions
can cause a disturbance to the values of the hidden variables of the system; a
possibility that is often overlooked in analyses of PPS paradoxes.
The paper is organized as follows. After introducing a general
form of the ABL rule, we give a simple example of how the
surprising features of a particular PPS paradox, known as the
three-box paradox, can be reproduced in a simple noncontextual
model that involves measurement-disturbance. Thereafter, we
consider the analogue of the ABL rule for HVTs, introduce the
assumption of noncontextuality, and introduce the notion of a
``logical'' PPS paradox. We then show how PPS paradoxes of this
type are consistent with noncontextuality if one allows for
measurement-disturbance. We end the paper with a brief discussion
of how a recent theorem that connects logical PPS
paradoxes to proofs of contextuality \citep{LeiSpek} appears in light of our
results.


\section{ Pre- and Post-Selection in Quantum Theory}

\label{PPSQT}

\subsection{Quantum Measurements}

We consider a finite dimensional Hilbert space and assume that no
evolution
occurs between measurements. The outcome of a quantum measurement, $\text{M}$%
, is a random variable, which we denote by $X_{\text{M}}$. We
restrict attention to the case where the range of $X_{\text{M
}}$is a discrete set labelled by $j.$

A measurement has both a \emph{statistical aspect}, which
specifies the probability of obtaining the different outcomes of
the measurement for any given density operator, and a
\emph{transformation aspect}, which specifies how the quantum
state is updated as a result of the measurement. We restrict our
attention to \emph{sharp} measurements, that is, those associated
with projector valued measures (PVMs) (sets of projectors
$\{P_{j}^{\text{M}}\}$ that sum to the identity operator,
$\sum_{j}P_{j}^{\text{M}}=I$). The probability of
obtaining outcome $X_{\text{M}}\ =j$ when the initial density matrix is $%
\rho $ is then given by $p_{\rho}(X_{\text{M}}\ =j )=\text{Tr}\left( P_{j}^{%
\text{M}}\rho \right) $. The most general possible transformation
aspect of
M is given by a set of completely positive (CP) maps $\{\mathcal{E}^\text{M}_j\}$ satisfying
\begin{equation}
\mathcal{E}^{\text{M}\;\#}_{j}(I)=P^{\text{M}}_{j}
\label{TransClass}
\end{equation}%
where $\mathcal{E}^{\#}$ is the dual of $\mathcal{E}$ defined by $\text{Tr}%
\left( \mathcal{E}^{\#}(A)B\right) =\text{Tr}\left( A\mathcal{E}(B)\right) $%
. The updated state on obtaining outcome $X_{\text{M}}\ =j$ is
\begin{equation}
\rho _{|X_{\text{M}}\ =j}=\frac{\mathcal{E}^{\text{M}}_{j}(\rho )}{\text{Tr}%
\left( \mathcal{E}^{\text{M}}_{j}(\rho )\right) }.
\label{UpdateRule}
\end{equation}%
When $\mathcal{E}^{\text{M}}_{j}(\rho )=P^{\text{M}}_{j}\rho
P^{\text{M}}_{j},$ the transformation is said to follow the
L\"{u}ders rule \citep{Luders}.

\subsection{Pre- and Post-Selected Systems}

Imagine an initial, an intermediate, and a final measurement
occurring at
times $t_{\text{pre}}$, $t$, and $t_{\text{post}}$ respectively, where $t_{%
\text{pre}}<t<t_{\text{post}}$. \ We pre-select (post-select) by
conditioning on a particular outcome of the initial (final)
measurement.
Denote the occurrence of this outcome by $\text{A}_{\text{pre}}$ ($\text{A}_{%
\text{post}}$), and suppose that it is associated with a projector $\Pi _{%
\text{pre}}$ $(\Pi _{\text{post}})$. Let the intermediate
measurement be denoted by $\text{M}$.

Assuming that the density operator prior to $t_{\text{pre}}$ is
$I/d,$ and assuming L\"{u}ders rule for the initial
measurement, the density operator after $t_{\text{pre}}$ is $\rho _{\text{pre%
}}=\Pi _{\text{pre}}/$Tr$(\Pi _{\text{pre}}).$ By Bayes' theorem,
we can deduce that the probability of obtaining the outcome $k$
for M is
\begin{equation}
p_{\text{ABL}}(X_{\text{M}}\ =k|\text{A}_{\text{pre}},\text{A}_{\text{post}},%
\text{M})=\frac{\text{Tr}(\Pi _{\text{post}}\mathcal{E}_{k}^{\text{M}}(\Pi _{%
\text{pre}}))}{\text{Tr}(\Pi _{\text{post}}\mathcal{E}_{k}^{\text{M}}(\Pi _{%
\text{pre}})+\Pi _{\text{post}}\mathcal{E}_{\lnot k}^{\text{M}}(\Pi _{\text{%
pre}}))}  \label{CoarseABL}
\end{equation}%
where $\mathcal{E}_{\lnot k}^{\text{M}}(\rho )=\sum_{j\neq k}\mathcal{E}^\text{M}_j(\rho )$. We refer to this as the ABL probability rule
\citep[see][]{AhaVaidReview}. From now on, unless stated otherwise, the
L{\"{u}}ders update rule is
assumed to apply for all intermediate measurements. \ In this case, $%
\mathcal{E}^{\text{M}}_{k}(\rho )=P^{\text{M}}_{k}\rho P^{\text{M}}_{k}$. \ Note that,
unlike the
standard Born rule probability, the ABL probability depends, through $%
\mathcal{E}_{\lnot k}^{\text{M}}$, on the entire PVM $\{P^{\text{M}}_{j}\}$
and not just on the projector $P^{\text{M}}_{k}$ that is associated with the
outcome for which the probability is being computed. \ This will
be critical further on.

\subsection{The Three-Box Paradox}

A simple example of the seemingly paradoxical predictions of the
ABL rule is the three-box paradox \citep[see][]{AhaVaidReview}. Suppose we
have a particle that can be in one of
three boxes. We label the state where the particle is in box $j$ by $%
\left\vert j\right\rangle $. The particle is pre-selected in the state $%
\left\vert \phi \right\rangle =\frac{1}{\sqrt{3}}\left( \left\vert
1\right\rangle +\left\vert 2\right\rangle +\left\vert
3\right\rangle \right) $, i.e. $\Pi _{\mathrm{pre}}=|\phi \rangle
\langle \phi |$, and
post-selected in the state $\left\vert \psi \right\rangle =\frac{1}{\sqrt{3}}%
\left( \left\vert 1\right\rangle +\left\vert 2\right\rangle
-\left\vert 3\right\rangle \right) $, i.e. $\Pi
_{\mathrm{post}}=|\psi \rangle \langle \psi |$. At an intermediate
time, we either determine whether the particle is in box $1$ or
not, or we determine whether it is in box $2$ or not. The
first measurement, $\text{M}$, corresponds to the PVM $\{P_{1}^{\text{M}%
},P_{2}^{\text{M}}\},$ where
\begin{equation}
P_{1}^{\text{M}}=\left\vert 1\right\rangle \left\langle
1\right\vert \qquad P_{2}^{\text{M}}=\left\vert 2\right\rangle
\left\langle 2\right\vert +\left\vert 3\right\rangle \left\langle
3\right\vert .
\end{equation}%
For this measurement, $p_{\text{ABL}}(X_{\text{M}}=1|\text{A}_{\text{pre}},\text{A}_{\text{post}},\text{M})=1$%
.

The second measurement, $\text{N}$, corresponds to the PVM $=\{P_{1}^{\text{N%
}},P_{2}^{\text{N}}\},$ where
\begin{equation}
P_{1}^{\text{N}}=\left\vert 2\right\rangle \left\langle
2\right\vert \qquad P_{2}^{\text{N}}=\left\vert 1\right\rangle
\left\langle 1\right\vert +\left\vert 3\right\rangle \left\langle
3\right\vert
\end{equation}%
In this case, $p_{\text{ABL}}(X_{\text{N}}=1|\text{A}_{\text{pre}},\text{A}_{\text{post}}
,\text{N})=1.$

Thus, if we ask whether or not the particle was in box $1$, we
find that it
was in box $1$ with certainty, and if we ask whether or not it was in box $2$%
, we find that it was in box $2$ with certainty!

\subsection{An Analogue of the three-box paradox}

\label{2Box}

Following the spirit of previous work by \citet{Kirk3Box}, we present a simple toy model that parallels the
features of the three-box paradox.

Consider an opaque box that can be divided into two opaque boxes
by placing a double partition in the box and breaking it into two
halves. It is also possible to put the two halves back together
and to remove the partition. The partition can be placed in two
possible positions, dividing the box either into front and back
halves or into right and left halves. Suppose there is a ball
inside the box. One can verify whether the ball is in the front
half of the box by dividing the box into front and back halves and
then shaking the front half of the box to hear whether the ball is
inside. If the ball is found in the front then this action
completely randomizes the left-right position of the ball.
However, if it is not found in the front then its position remains
undisturbed because the back half has not been shaken. A similar
procedure can be used to verify whether the ball is in the back
half of the box, only in this case the left-right position gets
randomized if the ball is in the back and is left undisturbed if
it is in the front. Two further procedures can be used to verify
whether the ball is in the left or right half of the box,
randomizing the front-back position of the ball whenever it is
found in the half that one is checking.

Now imagine that one pre-selects on finding the ball in the front
upon checking for it there, and one post-selects on finding the
ball in the back upon checking for it there. \ Suppose that the
two possible intermediate measurements are: 1) checking to see if
the ball is on the left, and 2) checking to see if the ball is on
the right. \ Clearly, if one checked to see if it was on the left,
then one must have found it on the left, since otherwise there
would have been no disturbance and consequently no way for the
ball to have moved from the front to the back of the box. \ But,
by the same argument, if one checked to see if it was on the
right, then one must have found it on the right.

This model succeeds at reproducing the surprising feature of the
three-box paradox, while being noncontextual according to the
operational definition provided in \citet{SpekkensContPMT}
(which we shall discuss further on). \ Moreover, it is clear the measurement-induced disturbance is critical
to achieving the surprising results.

In the rest of the paper, we show that this result is generic; PPS
paradoxes do not require contextuality for their explanation but
do require measurement-disturbance. The demonstration requires us
to examine explicitly how PPS scenarios are treated within an HVT.

\section{Hidden Variable Theories}

\label{HVT}

\subsection{Measurements in Hidden Variable Theories}

An HVT is an attempt to understand quantum measurements as
revealing features of \emph{ontic states}, by which we mean
complete specifications of the state of reality.  Let $\Omega $ be
the set of all ontic states in an HVT. Although $\lambda \in
\Omega $ are the states of reality, we may not have direct access
to them (hence the term \emph{hidden} variables), and so quantum mechanical preparation procedures
generally correspond to probability distributions over the ontic
states. We denote these by functions $\mu :$ $\Omega \rightarrow
\mathbb{R}_{+}$ satisfying $\int_{\Omega }\mu (\lambda )d\lambda
=1.$ \footnote{We assume that $\Omega$ satisfies the
measure-theoretic requirements necessary to make such integrals
well-defined.}

A common assumption that is made for HVTs is that the ontic state
determines the outcomes of all possible sharp measurements
uniquely. \ We refer to this assumption as \emph{outcome
determinism for sharp measurements}, and we presume it to hold in
all that follows. Given this assumption, the statistical aspect of
a measurement M is represented by a set of idempotent indicator
functions $\chi ^{\text{M}}_{j}:\Omega \rightarrow \{0,1\}$, that
sum to the unit function $\sum_{j}\chi ^{\text{M}}_{j}(\lambda
)=1$. $\chi ^{\text{M}}_{j}(\omega )$ is the probability ($0$ or
$1)$ of obtaining outcome $j$ given that the ontic state is
$\lambda $. More generally
\begin{equation}
p_{\mu}(X_{\text{M}}=j)=\int_{\Omega }\chi^\text{M}_j(\lambda )\mu (\lambda )d\lambda   \label{HVTProb}
\end{equation}%
is the probability of obtaining the outcome $j$ given the distribution $\mu $%
.

We now turn to the transformation aspect of measurements. We must
allow for the possibility that measurements cause a disturbance
(possibly stochastic) to the ontic state of the system. Therefore,
the disturbance induced by a measurement is described by a
transition matrix $D^{\text{M}}_{j}:\Omega
\times \Omega \rightarrow \mathbb{R}_{+}$ satisfying $\int_{\Omega }D^\text{M}_j(\lambda ,\omega )d\lambda =1.$ $D^{\text{M}}_{j}(\lambda
,\omega )$ is
the probability of a transition from $\omega $ to $\lambda $, given that $X_{%
\text{M}}\ =j$.

Thus, the most general transformation aspect of a measurement M
on obtaining outcome $j$ is composed of both a Bayesian updating of the
distribution and a disturbance, and is consequently represented by
a norm-decreasing
transition matrix%
\begin{equation}
\Gamma_{j}^{\text{M}}(\lambda ,\omega )=D_{j}^{\text{M}}(\lambda
,\omega )\chi _{j}^{\text{M}}(\omega ).  \label{decompofT}
\end{equation}%
This plays an analogous role in the HVT to the trace-decreasing CP
map associated with a measurement outcome in quantum theory. In
particular, it satisfies
\begin{equation}
\int_{\Omega }\Gamma_{j}^{\text{M}}(\lambda ,\omega )d\lambda
=\chi _{j}^{\text{M}}(\omega ),  \label{HVTdual}
\end{equation}%
which is the analogue of Eq.\ (\ref{TransClass}). We have the
following
update rule for the probability density on obtaining $X_{\text{M}}\ =j$%
\begin{equation}
\mu (\lambda |X_{\text{M}}=j)=\frac{\int_{\omega
}\Gamma^{\text{M}}_{j}(\lambda ,\omega )\mu (\omega )d\omega
}{\int_{\Omega }\Gamma^{\text{M}}_{j}(\lambda ,\omega )\mu (\omega
)d\omega d\lambda },  \label{HVTUpdate2}
\end{equation}%
which is the analogue of Eq.\ (\ref{UpdateRule}).

\subsection{Pre- and Post-Selection in Hidden Variable Theories}

Pre- and post-selected systems can be described in an HVT as
follows. The successful pre-selection event, A$_{\text{pre}}$, is
associated with a probability distribution $\mu
_{\text{pre}}(\lambda )$. The intermediate
measurement, $\text{M}$, is described by a set of indicator functions $%
\{\chi ^{\text{M}}_{j}\}$ and corresponding transition matrices $\{\Gamma_{j}^{%
\text{M}}\}$ and the successful post-selection event,
$\text{A}_{\text{post}} $ is associated with an indicator function
$\chi _{\text{post}}$. Applying Bayes' rule together with Eqs.\ (\ref{HVTProb}) and (\ref{HVTUpdate2}) we obtain the PPS
probability rule for HVTs
\begin{equation}
p_{\text{HVT}}(X_{\text{M}}=k|\text{A}_{\text{pre}},\text{A}_{\text{post}},%
\text{M})=\frac{\int_{\Omega }\chi _{\text{post}}(\lambda )\Gamma_{k}^{\text{M}%
}(\lambda ,\omega )\mu _{\text{pre}}(\omega )d\omega d\lambda
}{\int_{\Omega }\chi _{\text{post}}(\lambda
)(\Gamma_{k}^{\text{M}}(\lambda ,\omega )+\Gamma_{\lnot
k}^{\text{M}}(\lambda ,\omega ))\mu _{\text{pre}}(\omega )d\omega
d\lambda }. \label{PPSruleforHVT}
\end{equation}%
where $\Gamma^{\text{M}}_{\lnot k}(\lambda ,\omega )=\sum_{j\neq
k}\Gamma^{\text{M}}_{j}(\lambda ,\omega )$. This is the analogue
of Eq.\ (\ref{CoarseABL}).

\subsection{Measurement noncontextuality}

A particularly natural class of HVTs are the Measurement
Noncontextual Hidden Variable Theories (MNHVTs)
\citep[see][for a discussion of different types of noncontextuality]{SpekkensContPMT}.
The assumption of measurement noncontextuality
is that if there is an outcome $k$ of a
measurement $\text{M}$ and an outcome $j$ of a measurement $\text{N}$ that have the same probability for all possible
preparations, then they must be represented by the same indicator
function in the HVT. For quantum systems, this will be the case if
and only if both the outcomes correspond to the same projector
$P$, even though M and N may be associated with
different PVMs. Thus, noncontextuality implies that the indicator
functions in both cases depend only on the projector $P,$ i.e.
\begin{equation}
\chi^{\text{M}}_{k}(\lambda )=\chi _{j}^{\text{N}}(\lambda )=\chi _{P}(\lambda ). \label{MNC}
\end{equation}%
 Equivalently, in an MNHVT, every projector is associated
with a unique indicator function (defining a unique subset of
$\Omega$) which specifies the property that is revealed by a
measurement of that projector.

This implies stringent constraints on the probabilities that can
be simultaneously assigned to commuting projectors.  For instance,
since any pair of orthogonal projectors $Q_{1},Q_{2}$ can appear together in the same PVM, 
it follows that $\chi _{Q_{1}}+\chi
_{Q_{2}}\leq 1$ and consequently that $\chi _{Q_{1}}\chi
_{Q_{2}}=0$ (where we leave the dependence of $\chi $ on $\lambda
$ implicit). \ Moreover, if $Q=Q_{1}+Q_{2},$ then $Q$ is a
coarse-graining of $Q_{1}$ and $Q_{2},$ and therefore is
represented in the HVT by $\chi _{Q}=\chi _{Q_{1}}+\chi _{Q_{2}}.$
\ Given these identities, it follows that, for commuting projectors $P$
and $Q$, $\chi _{P}\chi _{Q}= \chi _{PQ}$ and $\chi _{P}+\chi _{Q}-\chi
_{P}\chi _{Q}= \chi _{P+Q-PQ}.$ \ Finally, the projector onto the null space, $P_{%
\text{null}},$ is represented by $\chi _{P_{\text{null}}}=0$ since
the associated property never holds. Denoting the probability one
assigns to $P$ given a distribution $\mu (\lambda )$, by
$p(P)=\int_{\Omega }\chi _{P}(\lambda )\mu (\lambda )d\lambda ,$
we obtain the following constraints

\textbf{Algebraic conditions: }For projectors $P,Q$ such that
$[P,Q]=0$,
\begin{align}
0& \leq p(P)\leq 1  \label{ac0} \\
p(I-P)& =1-p(P),  \label{ac1} \\
p(I)& =1,\;\text{ }p(P_{\text{null}})=0,  \label{ac2} \\
p(PQ)& \leq p(P),\;\text{ }p(PQ)\leq p(Q),  \label{ac3} \\
p(P+Q-PQ)& =p(P)+p(Q)-p(PQ).  \label{ac4}
\end{align}

If one conditions on a particular ontic state $\omega $, so that
$\mu (\lambda )=\delta (\lambda -\omega )$ and $p(P)=\chi
_{P}(\omega ),$ then all of these probabilities must be either $0$
or $1.$ \ However, the Bell-Kochen-Specker theorem shows that
there are sets of projectors for which there are no assignments of
probabilities $0$ or $1$ that satisfy the algebraic conditions.
This is an example of a proof of contextuality.

We are now in a position to make precise what it is about PPS
paradoxes that suggests that these might have something to do with
measurement contextuality. \ The critical fact is that there exist
sets of projectors that are each assigned probability $0$ or $1$
by the ABL rule such that the overall probability assignment
violates the algebraic conditions. The three
box paradox is an example of this. \ We call any such case a \emph{%
logical} PPS paradox.

First of all, let us clarify why, in the absence of the measurement noncontextuality assumption
, logical PPS paradoxes do not violate classical
probability theory. \ The reason is that only assignments that are
\emph{similarly conditioned} need to respect the algebraic
constraints, and when one conditions on the
nature of the intermediate measurement, the ABL probabilities \emph{do }%
satisfy the constraints. It is only if we consider ABL
probabilities for different intermediate measurements together that we find
that these probabilities do
not satisfy the functional relations defined in Eqs.\ (\ref{ac0}) to (\ref{ac4}%
). \ (For instance, we considered two distinct intermediate
measurements in the three box paradox.) \ Similarly, one can avoid
a contradiction in a HVT by assuming that indicator functions
depend on measurement context. In this case, a projector is not
associated with a unique indicator function and consequently can
be assigned different probabilities in different measurement
contexts.

This way of describing things suggests that logical PPS paradoxes
are themselves proofs of the impossibility of an MNHVT. \ This
only follows however if the dependence of the ABL probabilities on
context implies a similar dependence of the indicator functions on
context, equivalently, if context dependence of probabilities that
are conditioned on a pre- and post-selection of measurement
outcomes implies context dependence for probabilities that are
conditioned on a particular ontic state.

Since the HVT must reproduce the ABL probabilities, we can infer
that $p_{\text{HVT}}
(X_{\text{M}}=k|\text{A}_{\text{pre}},\text{A}_{\text{post}},\text{M})$
must be context-dependent. However, this does not immediately imply
context-dependence for the indicator functions.  One possible reason for this is that
$p_{\text{HVT}}(X_{\text{M}}=k|\text{A}_{\text{pre}},\text{A}_
{\text{post}},\text{M})$ depends on the transition matrices $\{\Gamma^{\text{M}}_j(\lambda
,\omega )\}$, and thus, given Eq.\ (\ref{decompofT}), it depends
not only on the indicator functions $\{\chi
_{j}^{\text{M}}(\lambda )\},$ but also on the transition matrices
for the disturbance, $\{D_{j}^{\text{M}}(\lambda ,\omega )\}.$ We
must therefore begin by analysing the possibility of context-dependence of the $D_{j}^{\text{M}}(\lambda
,\omega )$.

\subsection{Transformation noncontextuality}

In \citet{SpekkensContPMT}, the notion of noncontextuality is
generalized to transformations.  Transformation noncontextuality
is the assumption that equivalent transformations (i.e. those
represented by the same CP map in the quantum formalism) are
associated with the same transition matrix in an HVT. Thus, if
outcome $k$ of
a measurement $\text{M}$ and outcome $j$ of a measurement $\text{N}$ are both associated with the CP-map $\mathcal{E}$,
then
\begin{equation}
\Gamma_{k}^{\text{M}}(\lambda ,\omega )=\Gamma_{j}^{N}(\lambda ,\omega
)=\Gamma_{\mathcal{E}}(\lambda ,\omega ).
\end{equation}

If the measurement is sharp, that is, associated with a projector
$P,$ and the CP map $\mathcal{E}$ corresponds to the L\"{u}ders
rule, that is, $\mathcal{E}(\rho )=P\rho P,$ then a dependence on
$\mathcal{E}$ is simply a dependence on $P.$ \ In this case, the
assumption of transformation noncontextuality is that
$\Gamma_{k}^{\text{M}}(\lambda ,\omega )=\Gamma_{j}^{\text{N}}(\lambda ,\omega )=\Gamma_{P}(\lambda
,\omega )$. \ By Eq.\ (\ref{HVTdual}), it \ follows that $\chi
_{k}^{\text{M}}(\lambda )=\chi_j^\text{N}=\chi _{P}(\lambda ).$ Thus, for L\"{u}ders rule measurements,
transformation noncontextuality implies measurement
noncontextuality. \ So if we can
show that transformation noncontextuality is consistent with the
existence of logical PPS paradoxes, then we have also shown that
measurement noncontextuality is consistent with their existence.

\section{Main results}

As noted below eq.\ (\ref{CoarseABL}),
$p_{\text{ABL}}(X_{\text{M}
}=k|\text{A}_{\text{pre}},\text{A}_{\text{post}}, \text{M})$
depends on the identity of the entire PVM $\{P^\text{M}_{j}\}$ associated
with M and not just on the projector $P^\text{M}_{k}$ associated with the
outcome $k.$ \ Consequently, if the HVT is to reproduce the ABL
predictions, $p_{\text{HVT}}
(X_{\text{M}}=k|\text{A}_{\text{pre}},\text{A}_{\text{post}},\text{M})$
must also depend on $\{P^\text{M}_{j}\}$ and not just on $P^\text{M}_{k}$. \ We will
show that such a dependence can be achieved even under the
assumption of transformation noncontextuality.

First note that the
presence of $\Gamma_{\lnot k}^{\text{M}}(\lambda ,\omega )$ in Eq.\ (\ref{PPSruleforHVT}) shows that the PPS probability for HVTs \emph{can
}have a dependence on the PVM even under the assumption of
transformation noncontextuality, since $\Gamma_{\lnot
k}^{\text{M}}(\lambda ,\omega )=\sum_{j\neq
k}\Gamma_{j}^{\text{M}}(\lambda ,\omega )=\sum_{j\neq
k}\Gamma_{P^\text{M}_{j}}(\lambda ,\omega )$ depends on the entire PVM. We now show that it
\emph{must} have such a
dependence. Suppose M and N are associated with distinct PVMs, $%
\{P^\text{M}_{j}\}$ and $\{P_{j}^\text{N}\}$ where $P^{\text{M}}_{k}=P_{k}^\text{N} = P$ for some $%
k.$ \ Suppose further that the CP maps for each outcome are
described by the L\"{u}ders rule, so that the effective CP\ maps
associated with the \textquotedblleft not $k$\textquotedblright\
outcome are $\mathcal{E}_{\lnot
k}^{\text{M}}(\cdot )=\sum_{j\neq k}P^{\text{M}}_{j}(\cdot )P^{\text{M}}_{j}$ and $\mathcal{E}%
_{\lnot k}^{\text{N}}(\cdot )=\sum_{j\neq
k}P_{j}^{\text{N} }(\cdot )P_{j}^{\text{N} }.$ \ The distinctness of
$\{P_{j}^{\text{M}}\}$ and $\{P_{j}^{\text{N} }\}
$ then implies that $\mathcal{E}_{\lnot k}^{\text{M}}$ and $\mathcal{E}%
_{\lnot k}^{\text{N}}$ \ are distinct, which in turn
implies that there is some density operator $\rho$ that is mapped
to distinct density operators by $\mathcal{E}_{\lnot k}^{\text{M}}$ and $\mathcal{E}%
_{\lnot k}^{\text{N}}$.  Consequently, there must be
some distribution $\mu(\lambda)$ that is mapped to distinct
distributions by $\Gamma_{\lnot k}^{\text{M}}(\lambda ,\omega )$ and $\Gamma_{\lnot k}^{\text{N}}(\lambda ,\omega )$. 
However, this is only possible if $\Gamma_{\lnot k}^{\text{M}}(\lambda ,\omega )$ and $\Gamma_{\lnot k}^{\text{N}}(\lambda ,\omega )$ 
are themselves distinct transition
matrices.  Being distinct, they cannot depend only on $P,$ but
must depend on the full PVM. \ This concludes the proof.

 Finally, we prove that any transformation noncontextual HVT
that can reproduce the ABL predictions (of which logical PPS
paradoxes are an example) must involve measurement-disturbance.

We assume the contrary and derive a contradiction. If a
measurement involves no disturbance in a HVT, then
$\Gamma_{k}^{\text{M}}(\lambda ,\omega )=\delta (\lambda -\omega
)\chi _{k}^{\text{M}}(\omega )$ (which is simply Bayesian
updating). As noted above, $p_{\text{HVT}}(X_{\text{M}}=k|\text{A}_{\text{pre}},\text{A}_{\text{post}},\text{M})$ must
depend on
the full PVM rather than just $P^{\text{M}}_{k}$ to reproduce  $p_{\text{ABL}}(X_{\text{%
M}}=k|\text{A}_{\text{pre}},\text{A}_{\text{post}},\text{M})$.  Now, consider
the pair of measurements M and N introduced above. Since $\Gamma_{k}^{\text{M}}$ and
$\Gamma_{k}^{\text{N}}$ depend only on $P$ (by
virtue of transformation noncontextuality), it follows that
$\Gamma_{\lnot k}^{\text{M}}$ and $\Gamma_{\lnot
k}^{\text{N}}$ must
depend on the PVM. \ The absence of measurement-disturbance would imply $%
\Gamma_{\lnot k}^{\text{M}}(\lambda ,\omega )=\delta (\lambda
-\omega
)\sum_{j\neq k}\chi _{j}^{\text{M}}(\lambda )$ and $\Gamma_{\lnot k}^{\text{N}%
}=\delta (\lambda -\omega )\sum_{j\neq k}\chi
_{j}^{\text{N}}(\lambda ).$ \ But by the assumption of measurement noncontextuality, $%
\sum_{j\neq k}\chi _{j}^{\text{M}}(\lambda )=\sum_{j\neq k}\chi _{j}^{\text{N}}(\lambda )=\chi _{I-P}(\lambda ),$ which implies that $%
\Gamma_{\lnot k}^{\text{M}}(\lambda ,\omega )=\Gamma_{\lnot
k}^{\text{N}}(\lambda ,\omega ).$ However, if the
transition matrices are the same, then they do not depend on the
PVM context, and this contradicts the assumption that the HVT
reproduces the predictions of the ABL rule. \ This concludes the
proof.

\section{Outlook}

\label{Conc}

We have shown that the existence of logical PPS paradoxes in a
theory does not imply contextuality, which would seem to suggest that
the two phenomena are unrelated. However, we recently proved a
theorem \citep{LeiSpek} showing that the mathematical structure of
every logical PPS paradox in quantum mechanics is sufficient to
construct a proof of contextuality. This does not contradict the
results presented here because we did not show that a logical PPS
paradox is \emph{itself} a proof of contextuality; measurements that are
temporal successors in the PPS paradox must be treated as
counterfactual alternatives in the proof of contextuality.
This distinction is critical, since an earlier measurement can
cause a disturbance to the ontic state that is monitored by a
later measurement, whereas the possibility of having implemented a
different measurement cannot disturb the ontic state of the system
in the actual measurement. 

Nonetheless, there is some evidence
that within the framework of HVTs satisfying additional
constraints, one might find a closer connection between PPS
paradoxes and contextuality. For instance, in the model of \S
\ref{2Box} there is no analogue of the uncertainty principle,
since there is a nonzero probability of ascertaining both the
left-right and front-back position of the ball without disturbing
it in any way. \ Meanwhile, a toy theory that is noncontextual 
\citep[by the operational definition of][]{SpekkensContPMT} and that
\emph{does }include a strong analogue of the uncertainty
principle, that of \citet{SpekkensToy}, seems to be devoid of
logical PPS paradoxes. This suggests that there may be a natural
set of conditions that both quantum theory and the toy theory of
\citet{SpekkensToy} satisfy, but that the model of \S \ref{2Box}
and other HVTs do \emph{not} satisfy, under which contextuality
and logical PPS paradoxes are either both present or both absent.
\ We expect that some of the quantum structures discussed at this
conference might provide the appropriate setting to address this
question.

\bibliographystyle{plainnat}
\bibliography{contprepost}

\end{document}